\documentclass[prb,preprint]{revtex4}%
\usepackage{graphicx}
\usepackage{amsmath}
\usepackage{ulem}
\usepackage{color}

\newcommand{\be}{\begin{equation}}
\newcommand{\ee}{\end{equation}}
\newcommand{\bea}{\begin{eqnarray*}}
\newcommand{\eea}{\end{eqnarray*}}
\newcommand{\bean}{\begin{eqnarray}}
\newcommand{\eean}{\end{eqnarray}}

\begin{document}

\draft
\title{\bf Temperature-Stable Tunneling Current in Serial Double Quantum Dots: Insights from Nonequilibrium Green Functions and Pauli Spin Blockade}

\author{David M T Kuo}
%\DIFdelbegin \DIFdel{11
%}\DIFdelend

\address{Department of Electrical Engineering and Department of Physics, National Central
University, Chungli, 32001 Taiwan, Republic of China}

\date{\today}

\begin{abstract}
We theoretically investigate charge transport through serial
double quantum dots (SDQDs) with strong electron correlations
using nonequilibrium Green's function techniques. In the linear
response regime, we compute the charge stability diagram and
analyze the Coulomb oscillatory tunneling current, revealing both
thermal and nonthermal broadening effects on the current spectra
in relation to two gate voltages. In the nonlinear response
regime, we focus on tunneling currents in SDQDs under the Pauli
spin blockade (PSB) scenario. We find that current rectification
with negative differential conductance is significantly degraded
as temperature increases, making it challenging to distinguish
between the inter-site spin triplet and singlet states. Notably,
we observe a robust reversed tunneling current that remains stable
against temperature variations, provided the resonant channel in
the PSB scenario is coupled to the states of the right (left)
electrode, which is fully occupied (unoccupied) by particles. This
characteristic provides valuable insights for designing
transistors capable of operating over a wide temperature range.
\end{abstract}

\maketitle

\section{Introduction}
Impurities play a crucial role in influencing electron transport
in metals[\onlinecite{AndersonAW}-\onlinecite{GoldhaberG}], as
evidenced by phenomena such as Anderson
localization[\onlinecite{AndersonAW}-\onlinecite{YigalM}] and the
Kondo effect[\onlinecite{MadhavanV}-\onlinecite{GoldhaberG}].
These effects highlight the complex interactions that occur in
systems with impurities, necessitating advanced engineering
techniques to achieve precise control over electron transport at
the level of single atomic impurities. Notably, the resonant
electrical conductance peak exhibits distinctive temperature
behavior, which varies based on whether the temperature is below
or above the Kondo temperature. Expanding on the study of single
impurities, researchers are now investigating charge transport
through serial double quantum dots
(SDQDs)[\onlinecite{WaughFR}-\onlinecite{PettaJR}], which are
often referred to as artificial atoms. This area of research has
uncovered fascinating phenomena, including single-electron
coherent tunneling between quantum
dots[\onlinecite{WaughFR}-\onlinecite{Vanderw}] and current
rectification due to two-electron triplet state blockade
[\onlinecite{OnoK}-\onlinecite{Hendrickx}]. Furthermore, SDQDs
hold significant promise for applications in devices such as
quantum bits[\onlinecite{DiVincenzo}], quantum current sensors
[\onlinecite{Fujisawa}], and spintronics
[\onlinecite{Hanson},\onlinecite{ChenCC}]. However, a critical
challenge remains in ensuring reproducibility during the
implementation of SDQD devices.

Graphene nanoribbons (GNRs) present unique advantages due to the
topological states that emerge in the middle of their subband
structures. These states are well-isolated from both the
conduction and valence bands, allowing their wave functions to be
less susceptible to noise from other excited
modes[\onlinecite{GolorM}-\onlinecite{Mangnus}]. The bottom-up
synthesis technique provides precise control over GNR segments and
their heterojunctions at the atomic level. This level of precision
facilitates the formation of SDQDs without the typical issues
associated with size fluctuations and uncertainties in separation
distances[\onlinecite{DJRizzo}]. Moreover, the Coulomb
interactions, electron hopping strengths, and tunneling rates
within these topological states can be engineered to meet specific
requirements. When compared to double quantum dots constructed
from traditional materials like silicon, germanium, and gallium
arsenide[\onlinecite{Vanderw}-\onlinecite{Hendrickx}], the
reproducibility of GNR heterojunctions positions them as promising
building blocks for future quantum circuits[\onlinecite{BorsoiF}].

Despite significant theoretical advancements in understanding
charge transport through SDQDs within the Coulomb blockade
regime[\onlinecite{FranssonJ}-\onlinecite{Kondo}], deriving a
comprehensive formula for tunneling current that accounts for all
experimentally observed phenomena remains a
challenge[\onlinecite{WaughFR}-\onlinecite{Hendrickx}]. This study
presents an analytical expression for the transmission
coefficient, effectively describing the transport properties of
SDQDs. Specifically it focuses on SDQDs formed by the topological
states of 9-7-9 armchair graphene nanoribbons (AGNRs) in the
absence of spin-orbit interactions and external magnetic fields.
Figure 1 illustrates that the coupled localized interface states
of 9-7-9 AGNRs function as SDQDs[\onlinecite{ChenYC}]. Our
research focuses on elucidating the temperature dependence of
tunneling current in the Pauli spin blockade (PSB) configuration.
Notably, we find that nonthermal broadening results in reversed
tunneling currents that remain robust across a wide temperature
range. Transistors exhibiting these characteristics can make them
crucial components in circuits for artificial intelligence
applications or in extreme temperature environments, such as space
exploration, high-temperature industrial applications, and
scientific instruments used in harsh conditions.

\begin{figure}[h]
\centering
\includegraphics[trim=1.cm 0cm 1.cm 0cm,clip,angle=0,scale=0.3]{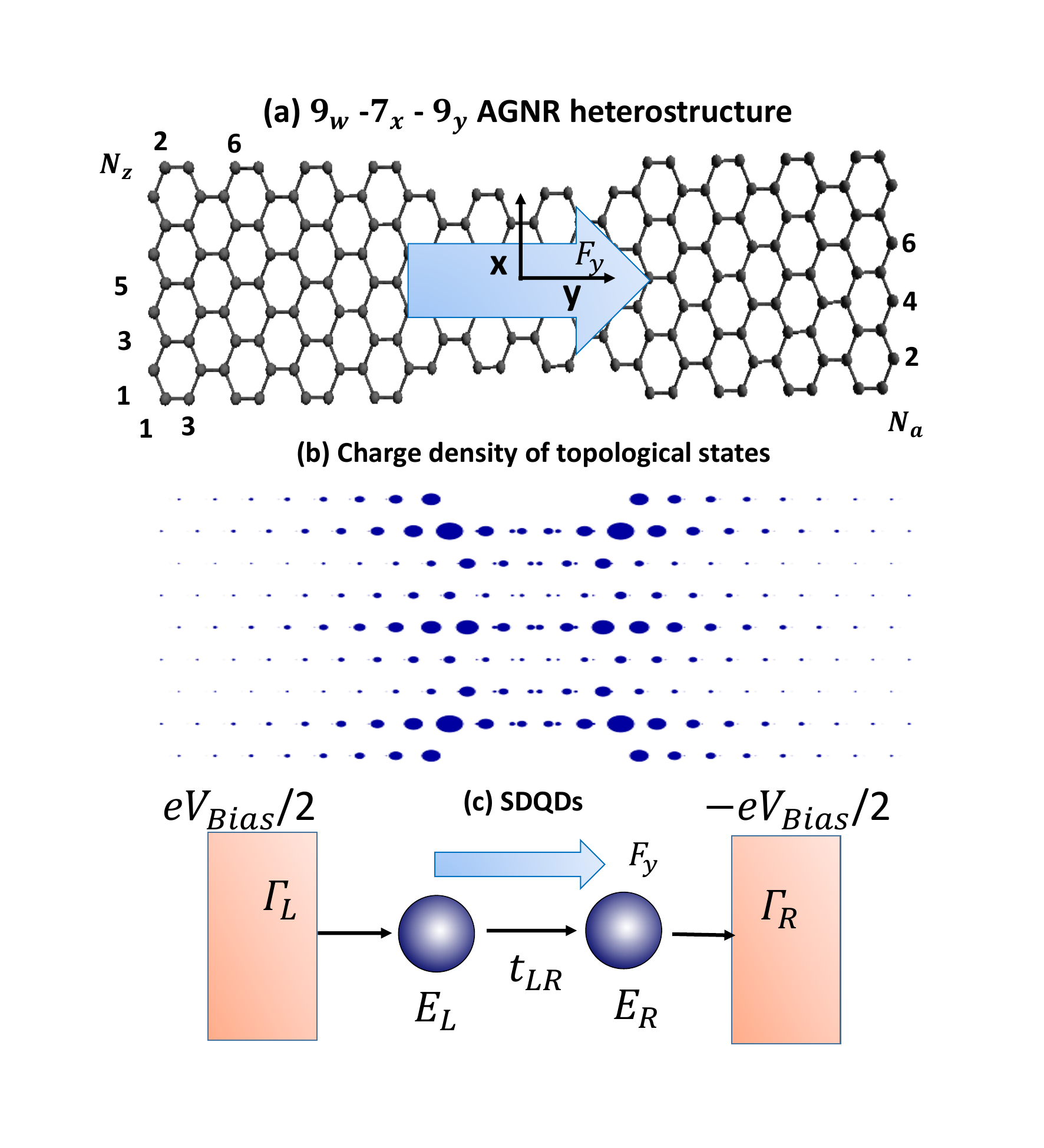}
\caption{(a) Schematic diagram of a finite $9_w-7_x-9_y$ armchair
graphene nanoribbon (AGNR) structure subjected to a uniform
electric field ($F_y$). The y-direction is defined along the
armchair edge structure of $9_w-7_x-9_y$ AGNR structure. The
subscripts w, x, and y in the notation $9_w-7_x-9_y$ denote the
segment lengths of the AGNR heterostructures in terms of the unit
cell (u.c.).(b)Charge density distribution of the topological
states in the $9_8-7_5-9_8$ AGNR sructure at $F_y = 0$. The radius
of each circle represents the magnitude of the charge density for
the topological states.(c) Charge transport through the interface
states of a 9-7-9 AGNR with zigzag edges, coupled to electrodes,
modeled as a serial double quantum dot system (SDQDs). The symbols
$\Gamma_{L}$ and $\Gamma_{R}$ denote the electron tunneling rates
between the left (right) electrode and the left (right) quantum
dot.
 }
\end{figure}

\section{Calculation Methodology}
Although many theoretical efforts have been devoted to studying
the electronic properties of AGNR heterojunctions
[\onlinecite{ChenYC}, \onlinecite{Kuo9}-\onlinecite{Kuo6}], the
electronic properties of 9-7-9 AGNR structures under a uniform
electric field, as illustrated in Fig. 1(a), remain unexplored.
The energy levels of $9_w-7_x-9_y$ AGNR structures under uniform
electric fields are presented in Appendix A, where we demonstrate
that the interface states are the sole channel for electron
transport through the topological states of the 9-7-9 AGNR
structure. These states effectively function as SDQDs, with each
quantum dot possessing only one energy level. To investigate
charge transport through SDQDs formed by AGNR heterojunctions in
the absence of spin-orbit interaction and magnetic fields, we
utilize a two-site Hubbard
model[\onlinecite{ChenYC},\onlinecite{Mangnus}]. The Hamiltonian
of this model, coupled to the electrodes, is expressed as $H =
H_0+H_{2-site}$, where

\begin{small}
\begin{eqnarray}
H_0& = &\sum_{k,\sigma} \epsilon_{k}
a^{\dagger}_{k,\sigma}a_{k,\sigma}+
\sum_{k,\sigma} \epsilon_{k} b^{\dagger}_{k,\sigma}b_{k,\sigma}\\
\nonumber &+& \sum_{k,\sigma}
V_{k,L}d^{\dagger}_{L,\sigma}a_{k,\sigma}
+\sum_{k,\sigma}V_{k,R}d^{\dagger}_{R,\sigma}b_{k,\sigma} + h.c.
\end{eqnarray}
\end{small}
The first two terms of Eq.~(1) describe the free electrons in the
left and right electrodes. $a^{\dagger}_{k,\sigma}$
($b^{\dagger}_{k,\sigma}$) creates an electron of momentum $k$ and
spin $\sigma$ with energy $\epsilon_k$ in the left (right)
electrode. $V_{k,L}$ ($V_{k,R}$) describes the coupling between
the left (right) lead and the left (right) site. Here,
$d^{\dagger}_{\ell,\sigma}$ ($d_{\ell,\sigma}$) creates (destroys)
an electron in the $\ell$ th site, where $\ell = L,R$ denotes the
left and right sites, respectively.

\begin{small}
\begin{eqnarray}
&& H_{2-site}\\ \nonumber &=&
\sum_{\ell,\sigma}E_{\ell}d^{\dagger}_{\ell,\sigma}d_{\ell,\sigma}
- t_{LR}~(d^{\dagger}_{R,\sigma}d_{L,\sigma} +
d^{\dagger}_{L,\sigma}d_{R,\sigma})\\ \nonumber & + &
\sum_{\ell}U_{\ell}~n_{\ell,\sigma}n_{\ell,-\sigma} +
\frac{1}{2}\sum_{j\neq\ell,\sigma,\sigma'}U_{j,\ell}~n_{j,\sigma}n_{\ell,\sigma'},
\end{eqnarray}
\end{small}

where $E_{\ell}$ represents the spin-independent energy level of
the localized state, and $n_{\ell,\sigma} =
d^{\dagger}_{\ell,\sigma}d_{\ell,\sigma}$. Notations $U_{\ell} =
U_{L(R)} = U_0$ and $U_{j,\ell} = U_{LR} = U_1$ denote the
intra-site and inter-site Coulomb interactions, respectively. For
simplicity, we set $U_0 = 150$~meV and $U_1 = 40$~meV for the
interface states of the $9_w-7_{10}-9_y$ AGNR
heterojunction[\onlinecite{Kuo9}]. These constant Coulomb
interaction values arise from the strong localization of the wave
functions of interface states, which remain largely unaffected by
the applied electric field within the considered range.

Using the equations of motion for the retarded and lesser Green's
functions [\onlinecite{JauhoAP},\onlinecite{Kuo8}], we derive a
closed-form expression for the tunneling current flowing from the
left (right) electrode, expressed as:

\begin{eqnarray}
& &J_{L(R)}(V_{bias},T)\\ \nonumber &=&\frac{2e}{h}\int
{d\varepsilon}~ {\cal
T}_{LR(RL)}(\varepsilon)[f_L(\varepsilon)-f_R(\varepsilon)].
\end{eqnarray}
The Fermi distribution function for the $\alpha$-th electrode is
defined as $f_{\alpha}(\varepsilon) =
1/(\exp[(\varepsilon-\mu_{\alpha})/k_BT]+1)$, where $\mu_{L(R)}=
E_F \pm eV_{bias}/2$ is the chemical potential of the left (right)
electrode with an applied bias of $V_{Bias}/2$ and $-V_{Bias}/2$,
respectively. The transmission coefficient ${\cal
T}_{LR}(\varepsilon)$ is expressed analytically as:

\begin{small}
\begin{eqnarray}
& &{\cal T}_{LR}(\epsilon)/(4t^2_{LR}\Gamma_{L}\Gamma_{R})=\frac{C_{1} }{|\epsilon_L\epsilon_R-t^2_{LR}|^2} \nonumber \\
&+& \frac{C_{2} }{|(\epsilon_L-U_{LR})(\epsilon_R-U_R)-t^2_{LR}|^2} \nonumber \\
&+& \frac{C_{3} }{|(\epsilon_L-U_{LR})(\epsilon_R-U_{LR})-t^2_{LR}|^2} \label{TF} \\
\nonumber &+&
\frac{C_{4} }{|(\epsilon_L-2U_{LR})(\epsilon_R-U_{LR}-U_R)-t^2_{LR}|^2}\\
\nonumber &+& \frac{C_{5} }{|(\epsilon_L-U_{L})(\epsilon_R-U_{LR})-t^2_{LR}|^2}\\
\nonumber &+& \frac{C_{6}
}{|(\epsilon_L-U_L-U_{LR})(\epsilon_R-U_R-U_{LR})-t^2_{LR}|^2}\\
\nonumber &+&
\frac{C_{7} }{|(\epsilon_L-U_L-U_{LR})(\epsilon_R-2U_{LR})-t^2_{LR}|^2}\\
\nonumber
 &+&
\frac{C_{8} }{|(\epsilon_L-U_L-2U_{LR})(\epsilon_R-U_R-2U_{LR})-t^2_{LR}|^2}, \\
\nonumber
\end{eqnarray}
\end{small}
where $\epsilon_L = \varepsilon-E_L+i\Gamma_{L}$ and $\epsilon_R =
\varepsilon-E_R+i\Gamma_{R}$. The tunneling rate
$\Gamma_{L(R)}(\varepsilon)= 2\pi \sum_{k,\sigma} |V_{k,L(R)}|^2
\delta(\varepsilon-\epsilon_k)$ arises from the QD coupled to
electrode. The transmission coefficient in Eq.(4) encompasses
eight terms, each corresponding to one of the eight possible
configurations of the SDQD that a particle with spin $\sigma$ from
the left electrode may encounter: (a) both sites empty, (b) one
particle with spin $-\sigma$ on the right site, (c) one particle
with spin $\sigma$ on the right site, (d) two particles on the
right-site, (e) one particle with spin $-\sigma$ on the left-site,
(f) both sites filled by one particle with spin $-\sigma$, (g) one
particle with spin $-\sigma$ on the left site and one particle
with spin $\sigma$ on the right site and (h) one particle with
spin $-\sigma$ on the left site and two particle on the
right-site. The probabilities of these configurations are
expressed as follow:

\begin{small}
\begin{eqnarray}
C_{1}&=&1-N_{L,\sigma}-N_{R,\sigma}-N_{R,-\sigma}+ \langle
n_{R,\sigma}n_{L,\sigma}\rangle \nonumber \\ &+&\langle
n_{R,-\sigma}n_{L,\sigma}\rangle+\langle
n_{R,-\sigma}n_{R,\sigma}\rangle-\langle
n_{R,-\sigma}n_{R,\sigma} n_{L,\sigma} \rangle \nonumber \\
C_{2}&=&N_{R,\sigma}-\langle n_{R,\sigma} n_{L,\sigma}\rangle
-\langle n_{R,-\sigma} n_{R,\sigma}\rangle \nonumber \\ &+&\langle
n_{R,-\sigma} n_{R,\sigma} n_{L,\sigma}\rangle \nonumber \\
C_{3}&=&N_{R,-\sigma}-\langle n_{R,-\sigma} n_{L,\sigma}\rangle
-\langle n_{R,-\sigma}n_{R,\sigma}\rangle \nonumber \\ &+&\langle
n_{R,-\sigma}n_{R,\sigma} n_{L,\sigma} \rangle \nonumber \\
C_{4}&=&\langle n_{R,-\sigma}n_{R,\sigma}\rangle-\langle
n_{R,-\sigma}n_{R,\sigma} n_{L,\sigma}\rangle \nonumber\\
C_{5}&=&N_{L,\sigma}- \langle n_{R,\sigma}n_{L,\sigma}\rangle
-\langle n_{R,-\sigma} n_{L,\sigma}\rangle \nonumber \\ &+&\langle
n_{R,-\sigma}n_{R,\sigma} n_{L,\sigma}\rangle \nonumber \\
C_{6}&=&\langle n_{R,\sigma} n_{L,\sigma}\rangle -\langle
n_{R,-\sigma}n_{R,\sigma} n_{L,\sigma}\rangle \nonumber \\
C_{7}&=&\langle n_{R,-\sigma}n_{L,\sigma}\rangle -\langle
n_{R,-\sigma}n_{R,\sigma} n_{L,\sigma}\rangle \nonumber \\
C_{8}&=&\langle n_{R,-\sigma}n_{R,\sigma}n_{L,\sigma} \rangle
\nonumber,
\end{eqnarray}
\end{small}
where $N_{\ell,\sigma}$ is the single particle occupation number
with spin $\sigma$ at site $\ell$. The intra-site and inter-site
two-particle correlation functions are denoted by $\langle
n_{\ell,-\sigma}n_{\ell,\sigma}\rangle$ and $\langle
n_{\ell,\sigma}n_{j,\sigma}\rangle$ ($\langle
n_{\ell,-\sigma}n_{j,\sigma}\rangle$), respectively. The
three-particle correlation function is represented by $\langle
n_{\ell,-\sigma}n_{\ell,\sigma}n_{j,\sigma}\rangle$. These
correlation functions can be solved self-consistently (see
Appendix B), ensuring that probability conservation is maintained,
as indicated by $\sum_{m}C_{m} = 1$. To obtain the reversed bias
tunneling current, we can simply exchange the indices of ${\cal
T}_{LR}(\varepsilon)$ in Eq. (4).

Note that the expression for the transmission coefficient in Eq.
(4) is valid only for temperatures above the Kondo
temperature[\onlinecite{MadhavanV}-\onlinecite{GoldhaberG}] and
for $U_1 > t_{LR}$[\onlinecite{Kuo8}]. This formulation of ${\cal
T}_{LR}(\varepsilon)$ extends beyond our previous work
[\onlinecite{DavidK}], where we provided the complete algebra for
the transmission coefficient and all correlation functions.
However, for computational simplicity, we only considered
one-particle occupation numbers and intra-site two-particle
correlation functions. This approximation omits spin-dependent
two-particle correlation functions, which are important for
electron transport under Pauli spin blockade (PSB) configurations
[\onlinecite{FranssonJ},\onlinecite{Inarrea}].

\section{Results and Discussion}
\subsection{Tunneling Current at Small Applied Bias}
To reflect the experimental conditions, we set the energies $E_L =
-(\delta_1eV_{g,L}+(1-\delta_1)eV_{g,R})+\eta~eV_{Bias}$ and $E_R
= -((1-\delta_2)eV_{g,L}+\delta_2V_{g,R})-\eta~eV_{Bias}$, where
$\delta_1$ and $\delta_2$ are determined by the inter-site and
intra-site electron Coulomb interactions[\onlinecite{SarmaSD}].
The parameter $\eta < 1/2 $ serves to adjust the orbital offset
induced by the applied bias $V_{Bias}$. $V_{g,L}$ and $V_{g,R}$
are the left and right gate voltages, respectively. In Fig. 2, we
present the calculated charge stability diagrams and tunneling
currents as functions of the two gate voltages for small applied
biases ($V_{Bias}$), aiming to replicate earlier experimental and
theoretical findings [\onlinecite{SarmaSD},\onlinecite{WangXin}],
where the authors considered the case of $\Gamma_{L(R)} = \Gamma_t
= 0$ and $ T = 0$.

As shown in Fig. 2(a), nine regions (plateaus) are marked by
integer values ($N_L, N_R$), where
$N_{L}=\sum_{\sigma}N_{L,\sigma}$ and
$N_{R}=\sum_{\sigma}N_{R,\sigma}$ represent the charge numbers of
the left and right sites, respectively. The total charge number
across the two sites is given by $N_t =
\sum_{\sigma}(N_{L,\sigma}+N_{R,\sigma})$. Since $N_{L(R)}$ is a
statistical average, $N_{L}$ and $N_{R}$ can take fractional
values in the transition areas of $N_t$, reflecting the
characteristics of an open system. The shapes of the regions
corresponding to ($N_L, N_R$) in the charge stability diagram
(Fig. 2(a)) are consistent with the results from Ref.
[\onlinecite{WangXin}], attributable to the conditions $k_BT <<
U_0$ and $t_{LR} \le \Gamma_t$. The charge stability diagram of
$t_{LR}
> \Gamma_t$ is illustrated in Appendix C.
Charge transport through graphene SDQD under the condition of $t_{LR}
> \Gamma_t$ (coherent tunneling) have been experimentally studied[\onlinecite{MolitorF}-\onlinecite{Banszerus}].
Our current study is limited to the case where $t_{LR} \le
\Gamma_t$.

Observable quantities, such as the spectra of electrical
conductance, provide deeper insights into the intricate structures
of SDQD devices, particularly regarding their contact properties
with electrodes. In Fig. 2(b), we present the tunneling current
corresponding to the charge stability diagram shown in Fig. 2(a).
When $V_{g,L} = V_{g,R}$, four peaks are identified, labeled
$P_1$, $P_3$, $P_6$ and $P_8$. The magnitudes of these peaks can
be analytically derived from the transmission coefficient given in
Eq. (4), and they are determined by their associated probability
weights $C_1$, $C_3$, $C_6$ and $C_8$. In contrast, $V_{g,L} \neq
V_{g,R}$, the peaks are labeled $P_2$, $P_4$, $P_5$ and $P_7$
 , corresponding to different channels with probability weights
$C_2$, $C_4$, $C_5$ and $C_7$ in Eq. (4). The results in Fig. 2(b)
indicate that the charge transfer process corresponds to a single
charge delocalization, as experimentally reported in Ref.
[\onlinecite{HatanoT}].

Additionally, references [\onlinecite{OnoK}-\onlinecite{Liruoyu}]
observed interesting tunneling current rectification even under
small applied biases. Consequently, we present the charge
stability diagram and tunneling current at the reversed bias
$V_{Bias}
 =-1 $~mV in Figs. 2(c) and 2(d). Due to the small applied bias
$V_{Bias}$, distinguishing the stability diagram in Fig. 2(c) from
that in Fig. 2(a) poses a challenge. However, notable differences
in the peaks labeled $P_2$ and $ P_4$ in Fig. 2(d) compared to
those in Fig. 2(b) are evident. These differences in magnitude
reflect the direction-dependent probabilities of tunneling current
through the channels associated with $C_2$ and $C_4$, aligning
well with the observations in references
[\onlinecite{OnoK}-\onlinecite{Liruoyu}].

\begin{figure}[h]
\centering
\includegraphics[trim=1.cm 0cm 1.cm 0cm,clip,angle=0,scale=0.3]{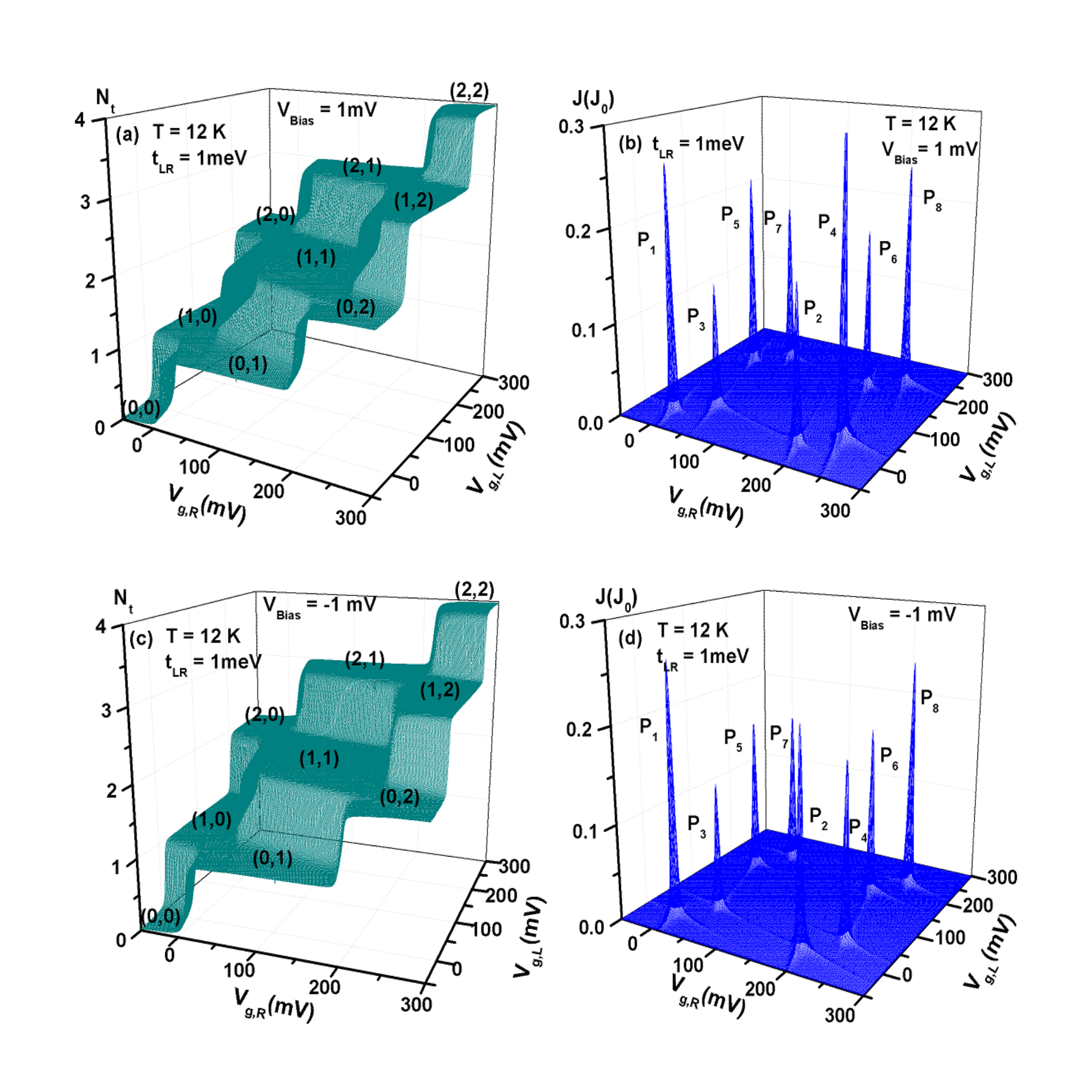}
\caption{(a)Total occupation number $N_t$ and (b) tunneling
current ($J = J_{L}$) as functions of gate voltages $V_{g,L}$ and
$V_{g,R}$ at a temperature of $T = 12$~K and a small applied bias
of $V_{Bias} = 1$~mV. (c) Total occupation number $N_t$ and (d)
tunneling current $J = |J_{R}|$ are calculated in the reversed
bias $V_{Bias}= -1$~mV and the same temperature of $T = 12$~K. The
analysis considers a AGNR heterojunction with $U_0 = 150$~meV,
$U_1 = 40$~meV and $\Gamma_L = \Gamma_R = \Gamma_t = 1$~meV. For
simplicity, we set $E_F = 0$, $\delta_1 = \delta_2 = 0.8$ through
out this article. The tunneling currents are expressed in units of
$J_0 = 0.773$~nA.}
\end{figure}

Next, we investigate the effect of temperature on the tunneling
current for both symmetrical  $V_{g,L} = V_{g,R} = V_g$ and
asymmetrical $ V_{g,R} \neq V_{g,L} $ scenarios. In Figs. 3(a) and
3(b), we present the tunneling currents as functions of $V_{g}$
and temperature $T$ at forward and reversed biases, respectively.
In Figs. 3(c) and 3(d), for the asymmetrical scenario, we show the
tunneling current as a function of $V_{g,R}$ and temperature $T$
with a fixed $eV_{g,L}= \frac{|4U_1-U_0|}{3}$, also at forward and
reversed biases. As illustrated in Figs. 3(a) and 3(b), the
thermal broadening of the tunneling currents decreases with
increasing temperature. This Coulomb oscillatory behavior has been
reported both experimentally and theoretically in Ref.
[\onlinecite{YigalM}]. In the symmetrical case, the charge
transport behavior resembles that of a single quantum dot with
multiple energy levels, and the current spectra are independent of
the direction of the applied bias $V_{Bias}$. Conversely, Figs.
3(c) and 3(d) reveal the presence of nonthermal broadening in the
tunneling currents. Here, the width of each tunneling current peak
exhibits a temperature-independent characteristic at high
temperatures, although its magnitude decays rapidly with
increasing temperature. The width of each peak can be approximated
as $2\sqrt{\Gamma_t^2+t^2_{LR}}$. Notably, the tunneling current
marked by $P_1$ in Fig. 3(d) differs from that in Fig. 3(c) at low
temperatures ($k_BT < 5$~meV). We observe an increase in tunneling
current with rising temperature in Fig. 3(d), which can be
attributed to the lower resonant level in the reversed bias
configuration. The nonthermal broadening characteristics evident
in Figs. 3(c) and 3(d) indicate that the asymmetrical scenario can
effectively filter thermionic charge transport arising from
temperature fluctuations.

\begin{figure}[h]
\centering
\includegraphics[angle=0,scale=0.3]{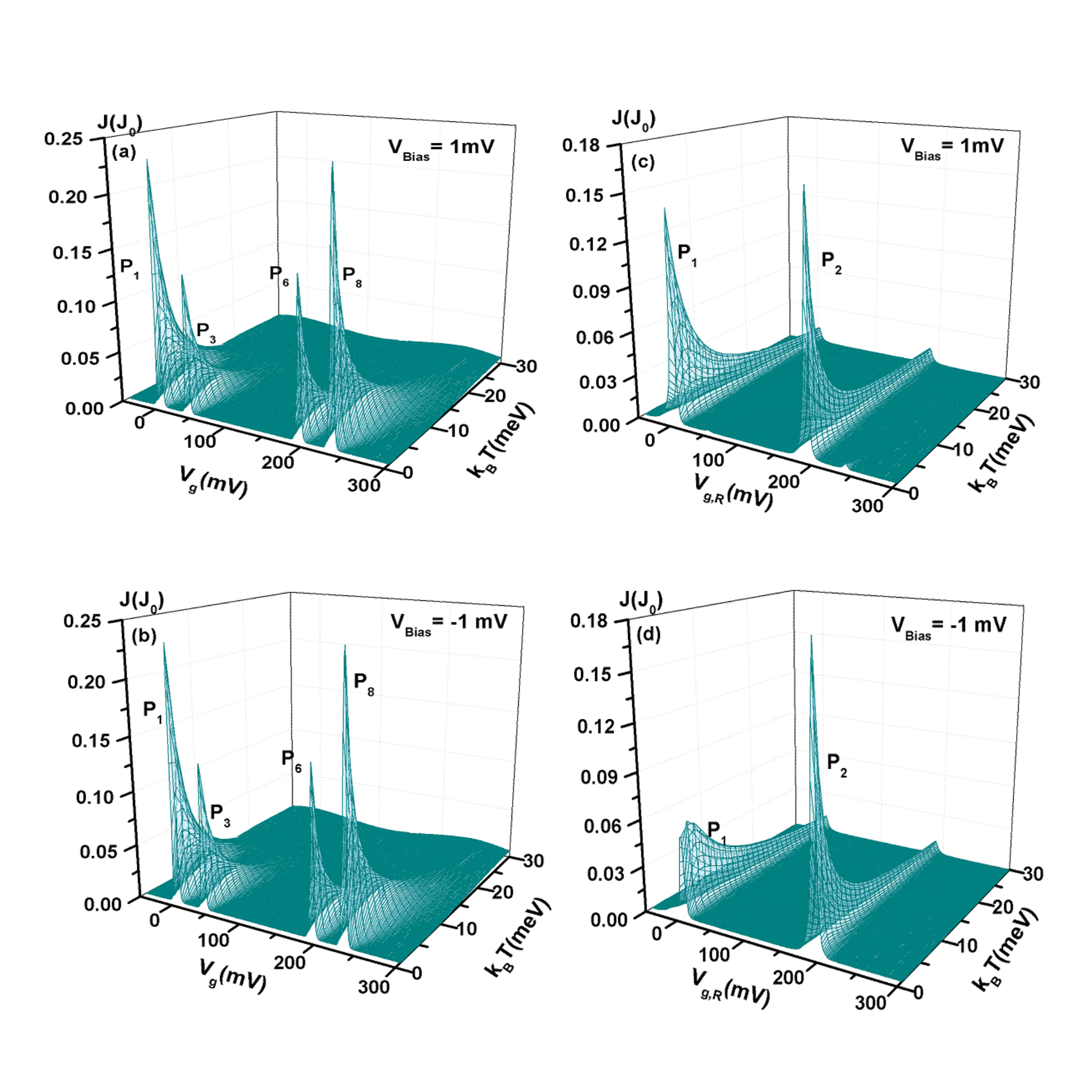}
\caption{Tunneling current as functions of $V_{g,L} = V_{g,R} =
V_g$ and $T$. (a) $V_{Bias}= 1$~mV and (b) $V_{Bias}= -1$~mV.
Tunneling current as functions of $V_{g,R}$ and $T$ at $eV_{g,L} =
|4U_1-U_0|/3$. (c) $V_{Bias}= 1$~mV and (d) $V_{Bias}= -1$~mV.
Other physical parameters are consistent with those in Fig.2.}
\end{figure}

\begin{figure}[h]
\centering
\includegraphics[angle=0,scale=0.3]{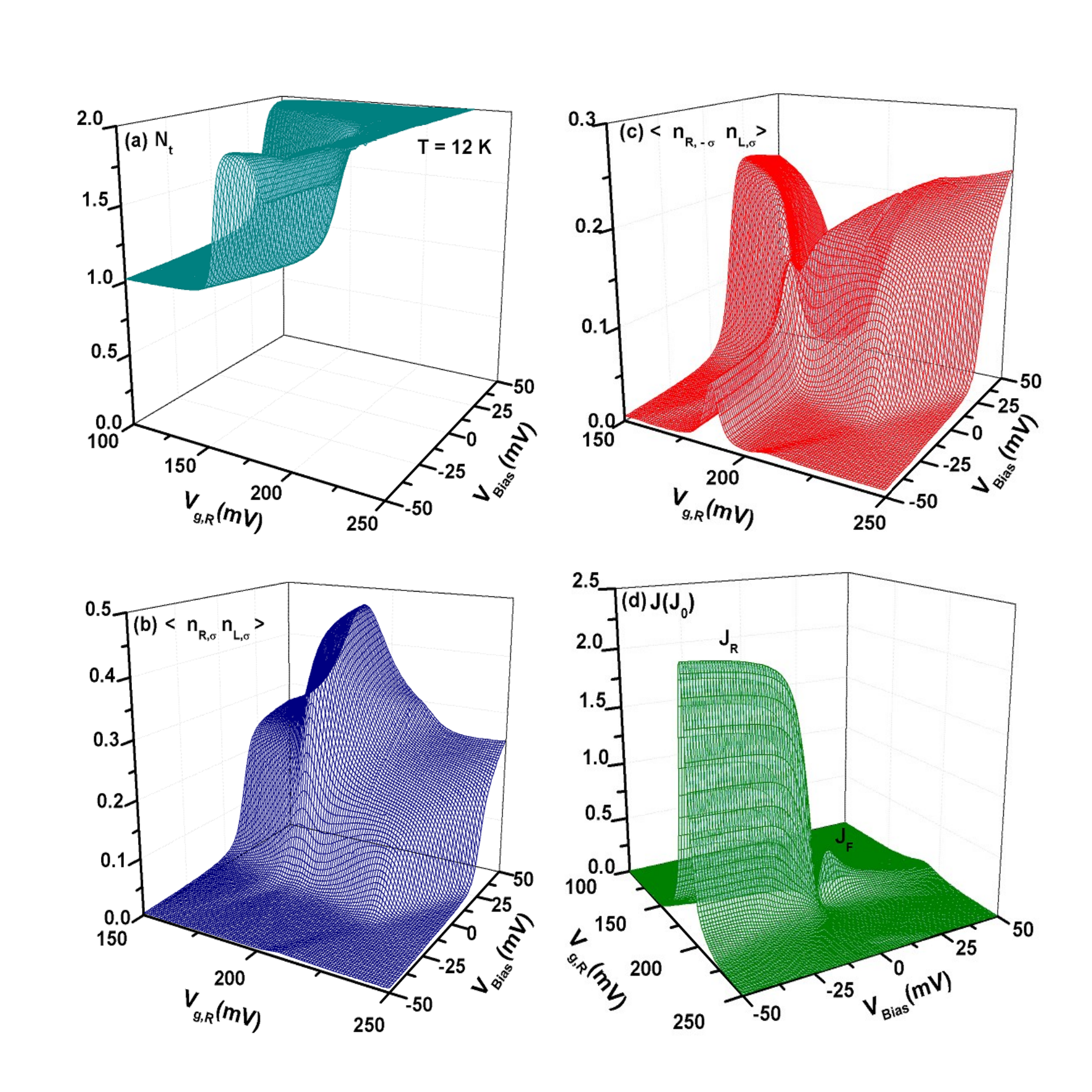}
\caption{(a)Total occupation number $N_t$, (b) inter-site
correlation function of triplet states $<  n_{R,\sigma}
n_{L,\sigma}
>$, (c) inter-site correlation function of singlet state $ <  n_{R,-\sigma} n_{L,\sigma}
>$ and (d) tunneling current as function of $V_{g,R}$ and $V_{Bias}$. The gate voltage for the left site is set to $eV_{g,L} =
(|4U_1-U_0|/3)$, with $ \eta = 0$ and $T = 12$~K. Other physical
parameters are consistent with those used in Fig.2.}
\end{figure}

\subsection{Tunneling Current at Large Bias}
In the two-site Pauli spin blockade (PSB) configuration, we set
$E_L + U_{1} = E_R + U_0 = E_F = 0 $ at $V_{bias} = 0$. This
condition can be achieved by tuning the gate voltages to $eV_{g,L}
= |4U_1-U_0|/3$ and $eV_{g,R} = |4U_0-U_1|/3$ (see $P_2$ in Fig.
2). To examine the charge transport of SDQDs under this PSB
configuration in the nonlinear response region, we present the
following quantities in Fig. 4: (a) total occupation number $N_t$,
(b) two-site correlation function of the triplet state $<
n_{R,\sigma}n_{L,\sigma}
> $, (c) two-site correlation function of the singlet state $<
n_{R,-\sigma} n_{L,\sigma}
>$ and (d) tunneling current $J$ as functions of $V_{g,R}$ and
$V_{Bias}$ at fixed $eV_{g,L} = |4U_1-U_0|/3$, $T = 12 $~K and
$\eta = 0$.

As shown in Fig. 4(a), $N_t$ is constrained within the range of $1
\le N_t \le 2$. At zero bias ($V_{Bias} = 0$), the values of the
inter-site correlation functions for the triplet state  $ <
n_{R,\sigma} n_{L,\sigma}>$ (Fig. 4(b)) and the singlet state $<
n_{R,-\sigma} n_{L,\sigma}
>$ (Fig. 4(c)) are difficult to distinguish. When a forward bias is
applied ($V_{Bias}
> 0$), the inter-site correlation function of the triplet state
significantly increases to approximately 0.5 with rising
$V_{Bias}$, while the inter-site correlation function of the
singlet state experiences substantial suppression. In contrast,
under reversed bias conditions ($V_{Bias} < 0$), the inter-site
correlation function of the triplet state becomes negligible.
Although the inter-site correlation function of the singlet state
exceeds that of the triplet state, it still diminishes with
increasing negative $V_{Bias}$ due to the right site being nearly
filled with two particles.

A notable rectification of the tunneling current is observed in
Fig. 4(d). The suppression of forward tunneling current is
attributed to the occupancy of both sites by the inter-site
triplet state[\onlinecite{FranssonJ},\onlinecite{Inarrea}].
Conversely, the enhancement of reversed current primarily results
from the left site being unoccupied. This current rectification
can be understood through the probabilities ($C_{F,PSB}$ and
$C_{R,PSB}$) associated with the forward and reversed tunneling
currents, given by:

\begin{small}
\begin{eqnarray}
C_{F,PSB} &=&N_{R,\sigma} - < n_{R,\sigma} n_{L,\sigma }> - <
n_{R,-\sigma} n_{R,\sigma }> \nonumber\\ &+ &< n_{R,-\sigma}
n_{R,\sigma } n_{L,\sigma}>
\end{eqnarray}
\end{small}

and

\begin{small}
\begin{eqnarray}
C_{R,PSB} &= &N_{R,\sigma} - < n_{L,\sigma } n_{R,\sigma}> - <
n_{L,-\sigma } n_{R,\sigma}
> \nonumber \\&+ &< n_{L,-\sigma} n_{L,\sigma } n_{R,\sigma}>
\end{eqnarray}
\end{small}

For the SDQD under PSB with $1 \le N_t \le 2$, we find the
probabilities of the $J_{F}$ ($J_R$) as $C_{F,PSB}= \frac{1}{2} -
< n_{R,\sigma} n_{L,\sigma }>$ and $C_{R,PSB} = 1 - < n_{L,\sigma
} n_{R,\sigma}> - < n_{L,-\sigma } n_{R,\sigma}
> $ [\onlinecite{JohnsonAC}]. This relationship effectively explains the current rectification
observed in Fig. 4.

The reversed tunneling current in Figure 4 does not exhibit the
characteristic of negative differential conductance (NDC), which
contrasts with experimental observations. In semiconductor SDQDs,
completely avoiding bias-induced orbital offsets is challenging.
To illustrate the $V_{Bias}$-dependent orbital offset effect
($\eta \neq 0$) (see appendix. A), we present the tunneling
current as functions of $V_{Bias}$ and $T$ under the PSB
configuration at $\eta = 0.1$ in Figure 5. In Figure 5(a), we now
observe NDC not only for forward applied bias but also for
reversed bias. The orbital offset causes a shift in the resonant
energy levels ($\varepsilon_L=E_L+U_1+\eta eV_{Bias} \neq
E_R+U_0-\eta eV_{Bias} =\varepsilon_R$), leading to reduce maximum
tunneling currents $J_{F,max}$ and $J_{R,max}$ compared to Figure
4(d) where $\eta = 0$. The tunneling current curves at low
temperature align well with experimental measurements
[\onlinecite{OnoK}, \onlinecite{JohnsonAC}].

As the temperature increases, the current rectification degrades,
and the NDC behavior of the tunneling current diminishes in Figure
5(a). According to Eqs. (5) and (6), the probabilities of charge
transport under the PSB configuration depend on single-particle
occupation numbers and particle correlation functions. Therefore,
we present the total occupation number ($N_t$), single-particle
occupation numbers ($N_{L,\sigma}$ and $N_{R,\sigma}$), and
inter-site spin correlation functions ($< n_{R,\sigma}
n_{L,\sigma} >$ and $< n_{R,-\sigma} n_{L,\sigma} >$) of the SDQDs
in Figures 5(b) through 5(f).

In Figure 5(b), the condition $1 \leq N_t \leq 2$ is maintained
over a wide temperature range, indicating that the values of the
three-particle correlation functions are small. As a result,
$C_{F,PSB}$ and $C_{R,PSB}$ are primarily determined by the
one-particle occupation number and two-particle correlation
functions. The behavior of $N_t$ shown in Figure 5(b) can be
further illustrated by the occupation numbers $N_{L,\sigma}$ and
$N_{R,\sigma}$ in Figures 5(c) and 5(d), respectively. Figures
5(c) and 5(d) show that, for $V_{Bias} > 0$, the two particles
tend to occupy different sites, whereas for $V_{Bias} < 0$ at low
temperatures ($k_B T = 1$~meV), they prefer to occupy the right
site together. This explains why $C_{F,PSB} = \frac{1}{2} - <
n_{R,\sigma} n_{L,\sigma} >$ and $C_{R,PSB} = 1 - < n_{L,\sigma}
n_{R,\sigma} > - < n_{L,-\sigma} n_{R,\sigma} >$ at low
temperatures ($k_B T = 1$~meV)[\onlinecite{Buitelaar}].

As the temperature increases beyond $10$~meV, we see that
$N_{L,\sigma}$ increases under $V_{Bias} < 0$, while
$N_{R,\sigma}$ is suppressed. At high temperatures ($k_B T >
10$~meV), $N_{L,\sigma}$ and $N_{R,\sigma}$ become less sensitive
to $V_{Bias}$. This behavior is reflected in the
temperature-dependent inter-site triplet and singlet correlation
functions shown in Figures 5(e) and 5(f). At high temperatures
($k_B T > 10$~meV), these two-particle correlation functions can
be approximated as the product of $N_{L,\sigma}$ and
$N_{R,\sigma}$. Compared to Figures 4(b) and 4(c), the orbital
offset reduces the distinction between the inter-site triplet and
singlet state curves. To effectively map two-electron spin states
via the tunneling current with NDC, it is crucial to minimize the
orbital offset effect and operate the SDQD at extremely low
temperatures.

\begin{figure}[h]
\centering
\includegraphics[angle=0,scale=0.3]{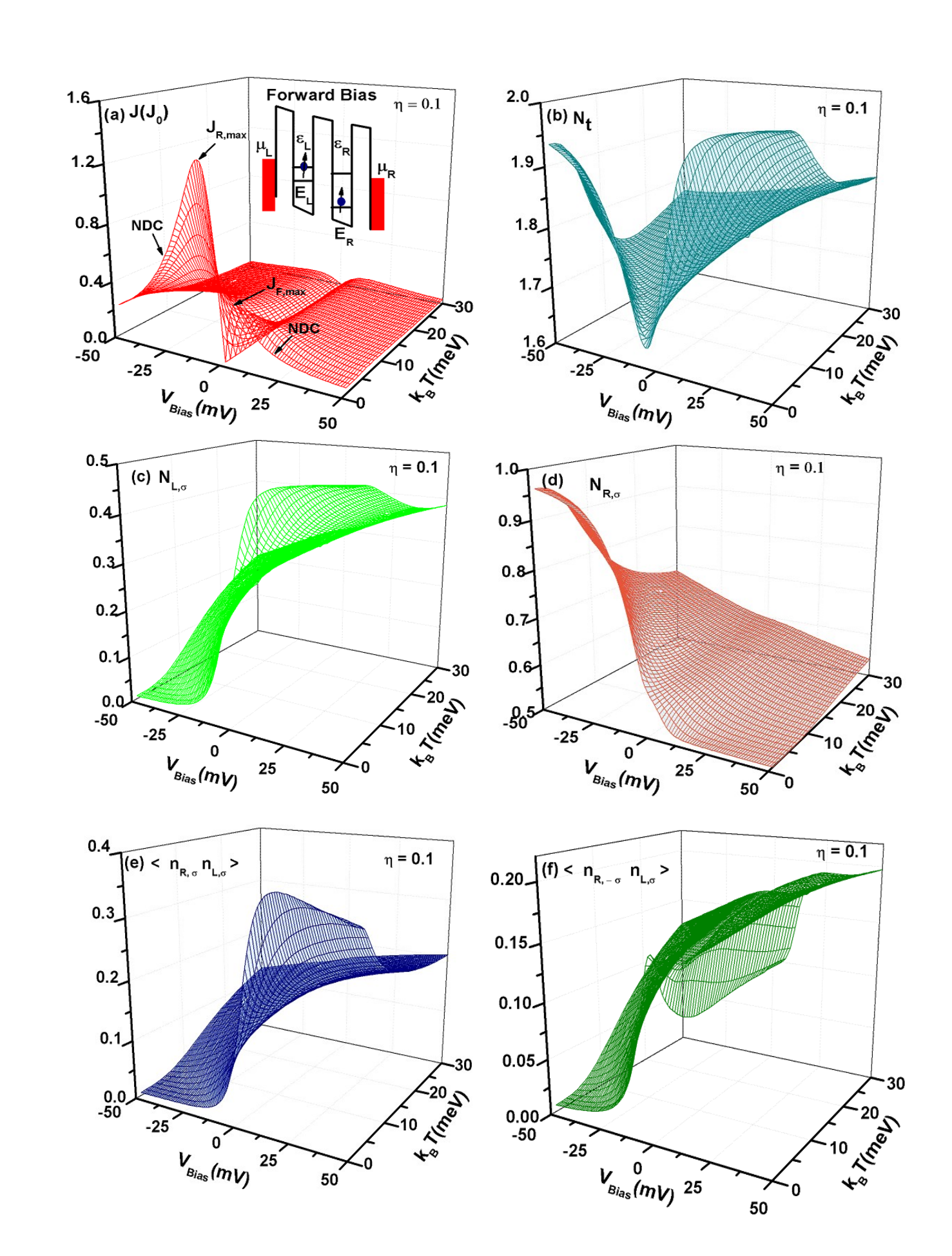}
\caption{(a) Tunneling current $J$, (b) total occupation number
$N_t$, (c) single particle occupation number of the left site
$N_{L,\sigma}$, (d) single particle occupation number of the right
site $N_{R,\sigma}$, (e) inter-site triplet correlation function
$< n_{R,\sigma} n_{L,\sigma} >$, and (f) inter-site singlet
correlation function $ < n_{R,-\sigma} n_{L,\sigma}> $ as
functions of $V_{Bias}$ and $T$. We have set $E_L = - U_1$, $E_R =
- U_0$, and $\eta = 0.1$. Other physical parameters are consistent
with those in Fig.2.}
\end{figure}

In Figure 5(a), the magnitude of tunneling currents significantly
decreases with increasing temperature. To achieve stable tunneling
currents across temperature variations, we consider the energy
levels $\Delta_L = E_L+U_1$ and $\Delta_R = E_R+U_0$ relative to
the Fermi energy ($E_F = 0$) of the electrodes. We present the
reversed tunneling currents as functions of $V_{Bias}$ and
temperature $T$ for various sets of $\Delta_L$ and $\Delta_R$ in
Figure 6 with $\eta = 0.3$. When $\Delta_L = 50$~meV and $\Delta_R
= -10$~meV, the maximum tunneling current occurs at $V_{Bias} = -
100$~mV as shown in Figure 6(a), resulting from $\varepsilon_L =
\Delta_L + \eta eV_{Bias} = \Delta_R - \eta eV_{Bias} =
\varepsilon_R$. However, the tunneling current decays rapidly with
increasing temperature. When $\Delta_L$ and $\Delta_R$ are
adjusted away from $E_F$, the resonant channel $\varepsilon_L =
\varepsilon_R$ is shifted to a larger reversed bias, resulting in
a tunneling current that is less sensitive to temperature. In
Figure 6(f), we observe a robust tunneling current even as the
temperature rises to room temperature. This temperature-stable
tunneling is attributed to the resonant channel $\varepsilon_L =
\varepsilon_R = 20$~meV being coupled to the states of the
electrodes under the conditions $\varepsilon_L-\mu_L \ge 2 k_BT$
and $\mu_R - \varepsilon_R \ge 2 k_BT$. The inset of Figure 6(f)
illustrates this context, where the Fermi distribution functions
indicate $f_R(\varepsilon_R)\approx 1$ and
$f_L(\varepsilon_L)\approx 0$.

\begin{figure}[h]
\centering
\includegraphics[angle=0,scale=0.3]{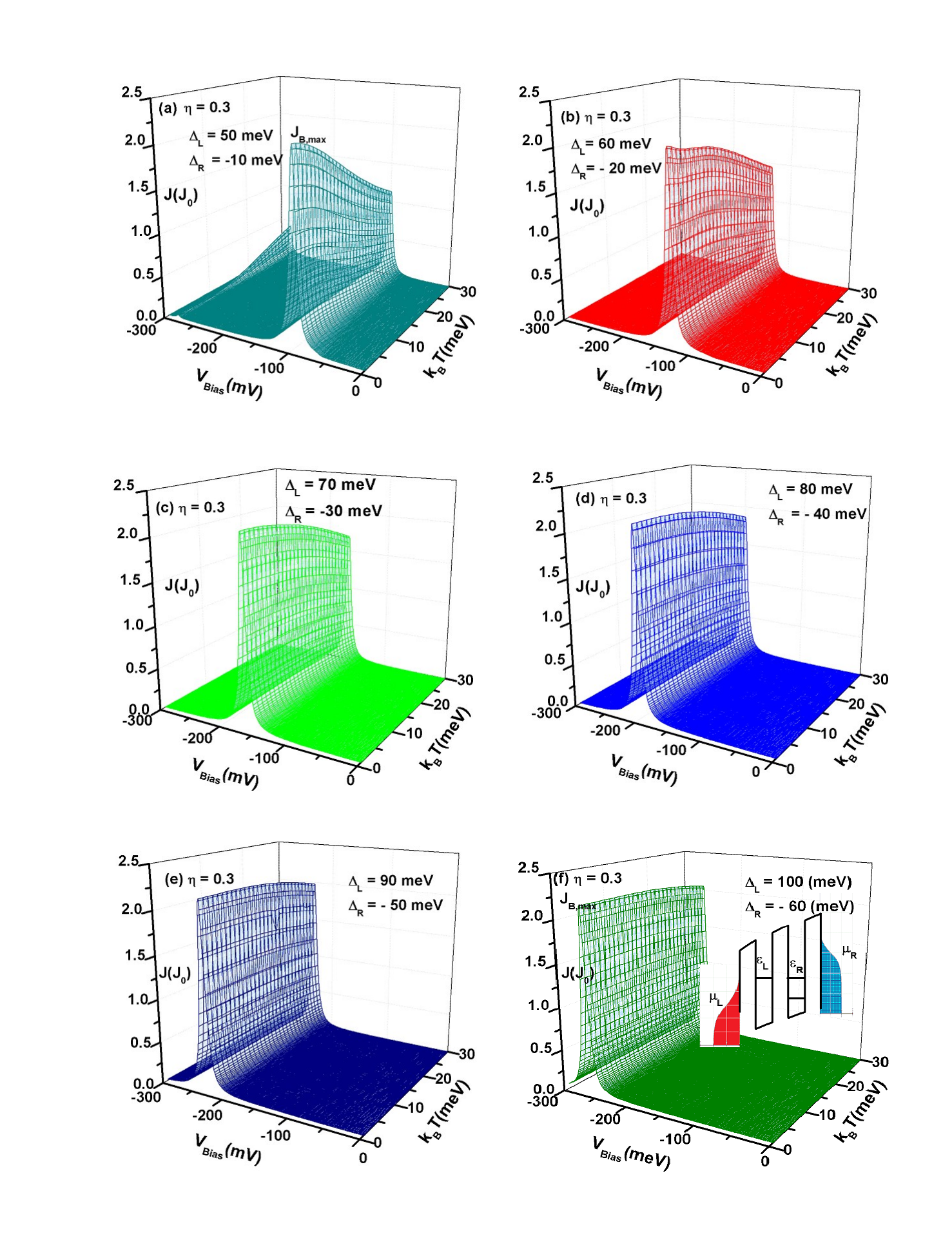}
\caption{Tunneling current as a function of $V_{Bias}$ and $T$ at
$\eta = 0.3$. The energy levels $\Delta_L = E_L + U_1$ and
$\Delta_R = E_R + U_0$ are introduced in the absence of applied
bias. (a) $\Delta_L = 50$~meV and $\Delta_R = -10$~meV, (b)
$\Delta_L = 60$~meV and $\Delta_R = -20$~meV, (c) $\Delta_L =
70$~meV and $\Delta_R = -30$~meV, (d) $\Delta_L = 80$~meV and
$\Delta_R = -40$~meV, (e) $\Delta_L = 90$~meV and $\Delta_R =
-50$~meV and (f) $\Delta_L = 100$~meV and $\Delta_R = -60$~meV. We
have considered $\eta = 0.3$. }
\end{figure}

\section{Conclusion}
The measurement of tunneling current through serial double quantum
dots (SDQDs) remains a vibrant area of research, as these
structures can be viewed as the smallest artificial molecules. The
tunneling current spectra of SDQDs offer insights into fundamental
physics and present promising applications for next-generation
quantum devices. While the physical principles governing tunneling
current spectra are understood, there is currently no unified
formula to explain key phenomena such as coherent tunneling,
Coulomb blockade, negative differential conductance (NDC),
nonthermal broadening of tunneling current, and current
rectification due to Pauli spin blockade.

Our study provides a closed-form expression for the tunneling
current, featuring an analytic transmission coefficient that
elucidates these experimental phenomena. This expression
incorporates one-particle occupation numbers, two-particle
inter-site correlation functions for both triplet and singlet
states, as well as intra-site and three-particle correlation
functions, all of which also have closed-form representations.
Consequently, the tunneling current of SDQDs can be easily
computed using general-purpose software such as Fortran, Python,
or Mathematica. This accessibility will assist designers
interested in implementing SDQD devices using various materials.

Using our closed-form tunneling current formula, we analyze the
tunneling current spectra through SDQDs formed by the topological
states of graphene nanoribbons. Beyond the major phenomena
mentioned, we introduce a novel application of SDQDs under a Pauli
spin blockade configuration, which allows for nonthermal
broadening of reversed tunneling currents that remain stable
across temperature variations. Transistors exhibiting
temperature-independent tunneling currents with NDC can function
effectively over a broad temperature range, making them crucial
components in circuits for artificial intelligence applications or
in extreme temperature environments.

%\begin{flushleft}

%\end{flushleft}
{}
%\begin{flushleft}

%\end{flushleft}

%\newpage
\mbox{}\\
\appendix{APPENDICES} \numberwithin{figure}{section}

\numberwithin{equation}{section}

\section{Stark shift of the topological states in a 9-7-9 AGNR
structure under an electric field}

To illustrate the Stark shift of the interface states in
$9_w-7_x-9_y$ AGNR structures under a uniform electric
field[\onlinecite{Kuo5},\onlinecite{Kuo6}], where w, x, and y
represent the number of unit cells in the AGNRs, we plot the
energy levels as functions of the applied bias $V_{Bias}$ for
various $x$ values in Fig. A. 1, with $w = y = 8$. In Fig. A.
1(a), notations $E_c$ and $E_v$ denote the energy levels for the
conduction and valence subband edge states, respectively. Symbols
$\Sigma_L$ and $\Sigma_R$ represent the left and right zigzag edge
states of $9_8-7_5-9_8$ AGNR structure, while $E_{I,L}$ and
$E_{I,R}$ correspond to the energy levels of the interface states
of the AGNR heterojunctions.

At zero applied bias, $\Sigma_L$ and $\Sigma_R$ nearly merge due
to the very weak coupling between the left and right zigzag edge
states. In contrast, $E_{I,L}$ and $E_{I,R}$ are separated by
$|2t_{LR}|$, where the electron hopping strength $t_{LR} =
37.7$~meV arising from the strong coupling between the left and
right interface states. The charge density for the interface
states of the $9_8-7_5-9_8$ AGNR structures is shown in Fig. 1(b).
The results in Fig. A. 1(a) indicate that while the subband states
($E_{c}$ and $E_v$) exhibit a red Stark shift, the topological
states ($\Sigma_L$, $\Sigma_R$, $E_{I,L}$ and $E_{I,R}$) show a
blue Stark shift (a linear function of $\eta~V_{bias}$), where
$\eta$ presents the slope of curve. When $V_{Bias}$ is smaller
than 0.5V, $E_{I,L}$ and $E_{I,R}$ remain well separated from the
subband states. As $x$ increases from $5$ to $10$, $t_{LR}$
becomes very small, resulting in $E_{I,L} = -E_{I,R}= t_{LR}
\approx 2.5$ meV for $x = 10$. Additionally, the parameter $\eta
\approx 0.215$ can be determined based on the behavior of
$E_{I,L(R)}$ as functions of $V_{Bias}$. The results in Fig. A. 1
demonstrate that $t_{LR}$ and $\eta$ can be engineered by varying
the configurations of 9-atom and 7-atom AGNR segments. Finally, it
is worth noting that the topological states $\Sigma_{L}$ and
$\Sigma_{R}$ do not exist in the transmission spectra of 9-7-9
AGNR with zigzag edges coupled to
electrodes[\onlinecite{Kuo5},\onlinecite{Kuo6}]. This highlights
that the interface states serve as the only channels for electron
tunneling through the topological states of 9-7-9 AGNR structures.

\begin{figure}[h]
\centering
\includegraphics[trim=1.cm 0cm 1.cm 0cm,clip,angle=0,scale=0.3]{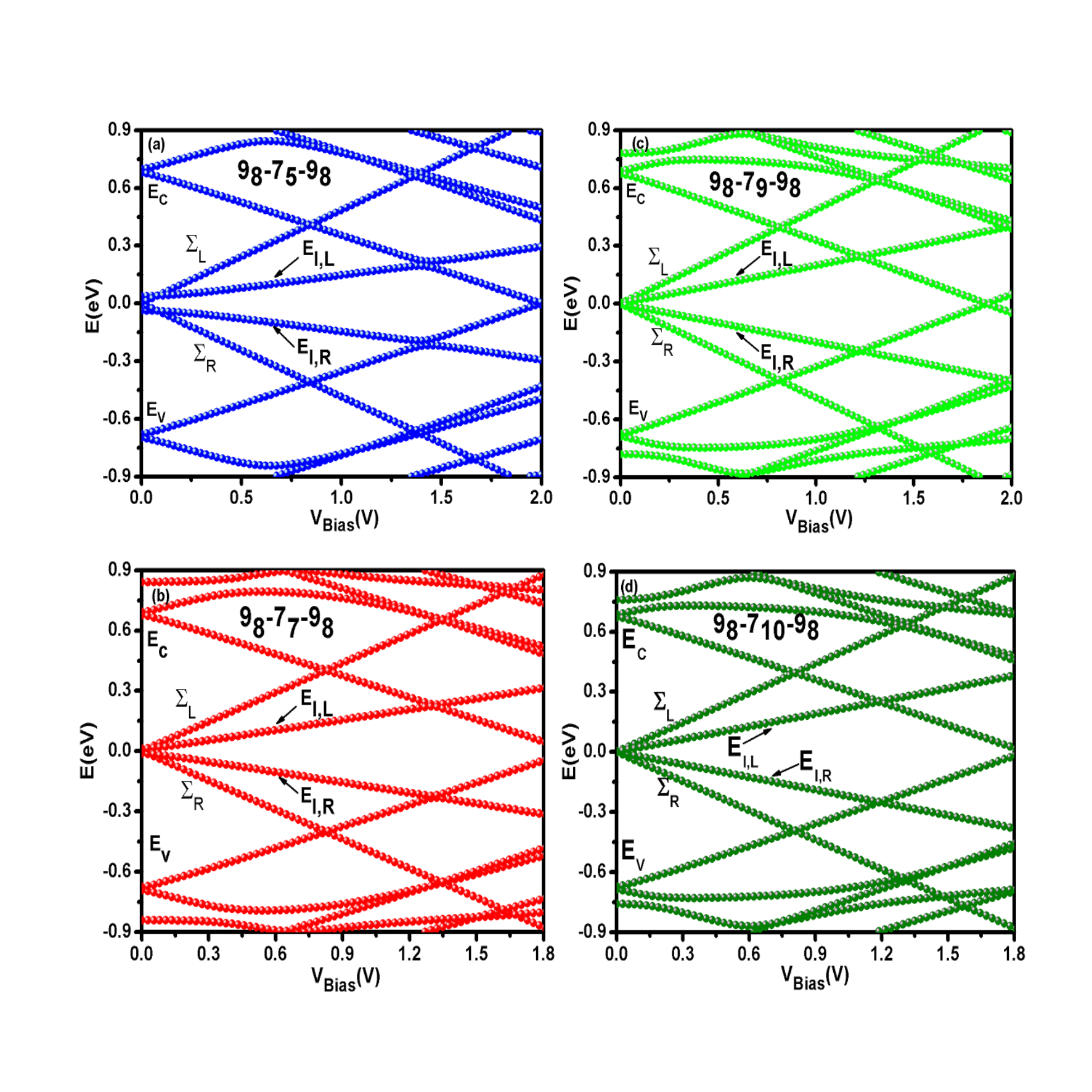}
\caption{Energy levels of 9-7-9 AGNR structure as functions of the
applied bias $V_{Bias}$ for various 7-atom AGNR segment lengths:
(a) $9_8-7_5-9_8$, (b) $9_8-7_7-9_8$, (c) $9_8-7_9-9_8$, and (d)
$9_8-7_{10}-9_8$.}
\end{figure}

%\numberwithin{figure}{section}

\section{Correlation functions of SDQDs}
%\numberwithin{figure}{section}

\numberwithin{equation}{section}

%\section{Appendix}~\Roman{section}
%\setcounter{section}{0}
 %\section{Appendix}
%\setcounter{equation}{0} % reset counter

The transmission coefficient of Eq. (4) describes a particle with
spin $\sigma$ tunneling through serial double quantum dots (SDQDs)
with energy levels $E_L$ and $E_R$, modulated by gate voltages and
the applied bias, respectively. Due to intra-site and inter-site
Coulomb interactions, we identify eight distinct configurations
with varying transport probabilities, which involve
single-particle occupation numbers, two-particle correlation
functions, and three-particle correlation
functions[\onlinecite{Kuo8}]. The closed-form expressions for
these quantities are provided below:

\textbf{a:One-Particle Occupation Number}

The one particle occupation number is given by
$N_{\ell,\sigma}=\int \frac{d\epsilon}{\pi}
G^<_{\ell,\sigma}(\epsilon)$, where the less Green's function
takes following form:

\begin{small}
\begin{eqnarray}
&&G^<_{\ell,\sigma}(\epsilon)\nonumber\\
 &&= \Sigma^<_{\ell}\Big[\frac{C_1}
{|\epsilon_{\ell}-t^2_{\ell,j}/\epsilon_j|^2 }+\frac{C_2}
{|(\epsilon_{\ell}-U_{\ell,j})-t^2_{\ell,j}/(\epsilon_j-U_j)|^2}\nonumber\\
\nonumber
&&+\frac{C_3}{|(\epsilon_{\ell}-U_{\ell,j})-t^2_{\ell,j}/(\epsilon_j-U_{j,\ell})|^2}\\
\nonumber
&&+\frac{C_4}{|(\epsilon_{\ell}-2U_{\ell,j})-t^2_{\ell,j}/(\epsilon_j-U_j-U_{j,\ell})|^2}\\
\nonumber
 &&+ \frac{C_5}
{|(\epsilon_{\ell}-U_{\ell})-t^2_{\ell,j}/(\epsilon_j-U_{j,\ell})|^2}\\
\nonumber &&+\frac{C_6}
{|(\epsilon_{\ell}-U_{\ell}-U_{\ell,j})-t^2_{\ell,j}/(\epsilon_j-U_j-U_{j,\ell})|^2}\\
\nonumber
&&+\frac{C_7}{|(\epsilon_{\ell}-U_{\ell}-U_{\ell,j})-t^2_{\ell,j}/(\epsilon_j-2U_{j,\ell})|^2}\\
\nonumber
&&+\frac{C_8}{|(\epsilon_{\ell}-U_{\ell}-2U_{\ell,j})-t^2_{\ell,j}/(\epsilon_j-U_j-2U_{j,\ell})|^2}\Big]\\
\nonumber &&+t^2_{\ell,j}\Sigma^<_j\Big[\frac{C_1}
{|\epsilon_{\ell}\epsilon_j-t^2_{\ell,j}|^2 }+ \frac{C_2}
{|(\epsilon_{\ell}-U_{\ell,j})(\epsilon_j-U_j)-t^2_{\ell,j}|^2}\\
\nonumber
&&+\frac{C_3}{|(\epsilon_{\ell}-U_{\ell,j})(\epsilon_j-U_{j,\ell})-t^2_{\ell,j}|^2}\\
\nonumber
&&+\frac{C_4}{|(\epsilon_{\ell}-2U_{\ell,j})(\epsilon_j-U_j-U_{j,\ell})-t^2_{\ell,j}|^2}\\
\nonumber
 &&+ \frac{C_5}
{|(\epsilon_{\ell}-U_{\ell})(\epsilon_j-U_{j,\ell})-t^2_{\ell,j}|^2}\\
\nonumber &&+\frac{C_6}
{|(\epsilon_{\ell}-U_{\ell}-U_{\ell,j})(\epsilon_j-U_j-U_{j,\ell})-t^2_{\ell,j}|^2}\\
\nonumber
&&+\frac{C_7}{|(\epsilon_{\ell}-U_{\ell}-U_{\ell,j})(\epsilon_j-2U_{j,\ell})-t^2_{\ell,j}|^2}\\
&&+\frac{C_8}{|(\epsilon_{\ell}-U_{\ell}-2U_{\ell,j})(\epsilon_j-U_j-2U_{j,\ell})-t^2_{\ell,j}|^2}\Big].\nonumber\\
\label{eq:ap_lessgl}
\end{eqnarray}
\end{small}
Here, the lesser self-energies are defined as
$\Sigma^{<}_{\ell}=\Gamma_{\ell} f_{\ell}(\varepsilon)$ and
$\Sigma^{<}_{j}(\varepsilon)=\Gamma_{j}f_{j}(\varepsilon)$, where
$\Gamma_{\alpha}$ represents the tunneling rate and
$f_{\alpha}(\varepsilon)$ is the Fermi distribution function of
the $\alpha$-th electrode. For simplicity, we assume a wide-band
limit for the tunneling rate and consider it to be
energy-independent. The notation $\epsilon_{\ell} =
\varepsilon-E_{\ell}+i\Gamma_{\ell}$ is also used. The
coefficients $C_m$ denote the probability weights of each
configuration in the transmission coefficient from Eq. (4).
Notably, due to the absence of spin-orbit interactions and the
magnetic field in the system Hamiltonian, the single-particle
occupation number is spin-independent: $N_{\ell,\sigma} =
N_{\ell,-\sigma}$. $N_{\ell,\sigma}$ is influenced by both
electrodes.

\textbf{b:Intra-site Singlet States}

The two-particle correlation function for intra-site singlet
states is given by:
\begin{small}
\begin{eqnarray}
& &\langle n_{j,-\sigma}n_{j,\sigma}\rangle \nonumber \\
&=&\int \frac{d\epsilon}{\pi}(\Sigma^{<}_j \Big[
\frac{D_1}{|(\epsilon_j-U_j)-t^2_{\ell,j}/(\epsilon_{\ell}-U_{\ell,j})|^2} \nonumber \\
&+& \frac{D_2}{|(\epsilon_j-U_j-U_{\ell,j})-t^2_{\ell,j}/(\epsilon_{\ell}-U_{\ell}-U_{\ell,j})|^2} \nonumber \\
&+&\frac{D_3}{|(\epsilon_j-U_j-U_{\ell,j})-t^2_{\ell,j}/(\epsilon_{\ell}-2U_{\ell,j})|^2}\nonumber \\
&+&\frac{D_4}{|(\epsilon_j-U_j-2U_{\ell,j})-t^2_{\ell,j}/(\epsilon_{\ell}-U_{\ell}-2U_{\ell,j})|^2}\Big] \nonumber \\
&+&t^2_{\ell,j}\Sigma^{<}_{\ell}\Big[\frac{D_1}{|(\epsilon_j-U_j)(\epsilon_{\ell}-U_{\ell,j})-t^2_{\ell,j}|^2} \nonumber \\
&+&\frac{D_2}{|(\epsilon_j-U_j-U_{\ell,j})(\epsilon_{\ell}-U_{\ell}-U_{\ell,j})-t^2_{\ell,j}|^2} \nonumber \\
&+&\frac{D_3}{|(\epsilon_j-U_j-U_{\ell,j})(\epsilon_{\ell}-2U_{\ell,j})-t^2_{\ell,j}|^2}\nonumber \\
&+&\frac{D_4}{|(\epsilon_j-U_j-2U_{L,j})(\epsilon_{\ell}-U_{\ell}-2U_{\ell,j})-t^2_{\ell,j}|^2}\Big]), \nonumber \\
\end{eqnarray}
\end{small}
The probability weights are defined as follows: $D_1=N_{j,\sigma}
-\langle n_{\ell,-\sigma}n_{j,\sigma}\rangle - \langle
n_{\ell,\sigma}n_{j,\sigma}\rangle + \langle
n_{\ell,-\sigma}n_{\ell,\sigma}n_{j,\sigma}\rangle$, $D_2=\langle
n_{\ell,\sigma}n_{j,\sigma}\rangle -\langle
n_{\ell,-\sigma}n_{\ell,\sigma}n_{j,\sigma}\rangle$, $D_3=\langle
n_{\ell,-\sigma}n_{j,\sigma}\rangle -\langle
n_{\ell,-\sigma}n_{\ell,\sigma}n_{j,\sigma}\rangle$, and
$D_4=\langle n_{\ell,-\sigma}n_{\ell,\sigma}n_{j,\sigma}\rangle$.
In the Pauli spin configuration, $\langle
n_{j,-\sigma}n_{j,\sigma}\rangle $ is primarily influenced by the
term with  $D_1$.

\textbf{c:Inter-Site Triplet State}

The two-particle correlation function for inter-site triplet
states is expressed as:
\begin{small}
\begin{eqnarray}
& &\langle n_{j,\sigma}n_{\ell,\sigma}\rangle \nonumber \\
&=&\int \frac{d\epsilon}{\pi}(\Sigma^{<}_{\ell}\Big[\frac{F_1}{|(\epsilon_{\ell}-U_{\ell,j})-t^2_{\ell,j}/(\epsilon_j-U_{\ell,j})|^2} \nonumber \\
&+&\frac{F_2 }{|(\epsilon_{\ell}-U_{\ell}-U_{\ell,j})-t^2_{\ell,j}/(\epsilon_j-2U_{\ell,j})|^2} \nonumber \\
&+&\frac{F_3}{|(\epsilon_{\ell}-2U_{\ell,j})-t^2_{\ell,j}/(\epsilon_j-U_j-U_{\ell,j})|^2}\nonumber \\
&+&\frac{F_4}{|(\epsilon_{\ell}-U_{\ell}-2U_{\ell,j})-t^2_{\ell,j}/(\epsilon_j-U_j-2U_{\ell,j})|^2}\Big]\nonumber \\
&+&t^2_{\ell,j}\Sigma^{<}_j(\varepsilon)\Big[
\frac{F_1}{|(\epsilon_{\ell}-U_{\ell,j})(\epsilon_j-U_{\ell,j})-t^2_{\ell,j}|^2} \nonumber \\
&+&\frac{F_2 }{|(\epsilon_{\ell}-U_{\ell}-U_{\ell,j})(\epsilon_j-2U_{\ell,j})-t^2_{\ell,j}|^2} \nonumber \\
&+&\frac{F_3}{|(\epsilon_{\ell}-2U_{\ell,j})(\epsilon_j-U_j-U_{\ell,j})-t^2_{\ell,j}|^2}\nonumber \\
&+&\frac{F_4}{|(\epsilon_{\ell}-U_{\ell}-2U_{\ell,j})(\epsilon_j-U_j-2U_{\ell,j})-t^2_{\ell,j}|^2}\Big])\nonumber \\
\end{eqnarray}
\end{small}

where we have $F_1=N_{j,-\sigma} -\langle
n_{j,-\sigma}n_{\ell,\sigma}\rangle - \langle
n_{j,-\sigma}n_{j,\sigma}\rangle + \langle
n_{j,-\sigma}n_{j,\sigma}n_{\ell,\sigma}\rangle$, $F_2=\langle
n_{j,-\sigma}n_{\ell,\sigma}\rangle -\langle
n_{j,-\sigma}n_{j,\sigma}n_{\ell,\sigma}\rangle$, $F_3=\langle
n_{j,-\sigma}n_{j,\sigma}\rangle -\langle
n_{j,-\sigma}n_{j,\sigma}n_{\ell,\sigma}\rangle$, and $F_4=\langle
n_{j,-\sigma}n_{j,\sigma}n_{\ell,\sigma}\rangle$.

\textbf{d:Inter-Site Singlet State}

The two-particle correlation function for the inter-site singlet
state is expressed as:

\begin{small}
\begin{eqnarray}
& &\langle n_{j,-\sigma}n_{\ell,\sigma}\rangle \nonumber \\
&=&\int \frac{d\epsilon}{\pi}(\Sigma^{<}_{\ell}\Big[
\frac{G_{1}}{|(\epsilon_{\ell}-U_{\ell,j})-t^2_{\ell,j}/(\epsilon_j-U_{j})|^2} \nonumber \\
&+&\frac{G_2}{|(\epsilon_{\ell}-U_{\ell}-U_{\ell,j})-t^2_{\ell,j}/(\epsilon_j-U_j-U_{\ell,j})|^2} \nonumber \\
&+&\frac{G_3}{|(\epsilon_{\ell}-2U_{\ell,j})-t^2_{\ell,j}/(\epsilon_j-U_j-U_{\ell,j})|^2}\nonumber \\
&+&\frac{G_4}{|(\epsilon_{\ell}-U_{\ell}-2U_{\ell,j})-t^2_{\ell,j}/(\epsilon_j-U_j-2U_{\ell,j})|^2}\Big]\nonumber \\
&+&t^2_{\ell,j}\Sigma^{<}_j\Big[\frac{G_{1}}{|(\epsilon_{\ell}-U_{\ell,j})(\epsilon_j-U_{j})-t^2_{\ell,j}|^2} \nonumber \\
&+&\frac{G_2}{|(\epsilon_{\ell}-U_{\ell}-U_{\ell,j})(\epsilon_j-U_j-U_{\ell,j})-t^2_{\ell,j}|^2} \nonumber \\
&+&\frac{G_3}{|(\epsilon_{\ell}-2U_{\ell,j})(\epsilon_j-U_j-U_{\ell,j})-t^2_{\ell,j}|^2}\nonumber \\
&+&\frac{G_4}{|(\epsilon_{\ell}-U_{\ell}-2U_{\ell,j})(\epsilon_j-U_j-2U_{\ell,j})-t^2_{\ell,j}|^2}\Big]).\nonumber \\
\end{eqnarray}
\end{small}
The probability weights are defined as: $G_1=N_{j,\sigma} -\langle
n_{j,\sigma}n_{\ell,\sigma}\rangle - \langle
n_{j,-\sigma}n_{j,\sigma}\rangle + \langle
n_{j,-\sigma}n_{j,\sigma}n_{\ell,\sigma}\rangle$, $G_2=\langle
n_{j,\sigma}n_{\ell,\sigma}\rangle -\langle
n_{j,-\sigma}n_{j,\sigma}n_{\ell,\sigma}\rangle$, $G_3=\langle
n_{j,-\sigma}n_{j,\sigma}\rangle -\langle
n_{j,-\sigma}n_{j,\sigma}n_{\ell,\sigma}\rangle$, and $G_4=\langle
n_{j,-\sigma}n_{j,\sigma}n_{\ell,\sigma}\rangle$.

\textbf{e:Three Particle Correlation Functions }

The three-particle correlation function is given by:
\begin{small}
\begin{eqnarray}
& &\langle n_{j,-\sigma}n_{j,\sigma}n_{\ell,\sigma}\rangle \nonumber \\
&=&\int \frac{d\epsilon}{\pi}(\Sigma^{<}_{\ell} \Big[
\frac{W_1}{|(\epsilon_{\ell}-2U_{\ell,j})-t^2_{\ell,j}/(\epsilon_{j}-U_{j}-U_{\ell,j})|^2} \nonumber \\
&+&
\frac{W_2}{|(\epsilon_{\ell}-U_{\ell}-2U_{\ell,j})-t^2_{\ell,j}/(\epsilon_{j}-U_{j}-2U_{\ell,j})|^2}\Big] \nonumber \\
&+&t^2_{\ell,j}\Sigma^{<}_{j}\Big[\frac{W_1}{|(\epsilon_{\ell}-2U_{\ell,j})(\epsilon_{j}-U_{j}-U_{\ell,j})-t^2_{\ell,j}|^2} \nonumber \\
&+&\frac{W_2}{|(\epsilon_{\ell}-U_{\ell}-2U_{\ell,j})(\epsilon_{j}-U_{j}-2U_{\ell,j})-t^2_{\ell,j}|^2}
\Big]). \nonumber \\
\end{eqnarray}
\end{small}
The probability weights are defined as: $W_1=\langle
n_{j,-\sigma}n_{j,\sigma}\rangle -\langle
n_{j,-\sigma}n_{j,\sigma}n_{\ell,\sigma}\rangle$, and $W_2=\langle
n_{j,-\sigma}n_{j,\sigma}n_{\ell,\sigma}\rangle$. Using Eqs. (B.1)
to (B.5), we solve these correlation functions self-consistently.
Once we obtain the solutions, we substitute them back into the
transmission coefficient from Eq. (4) to calculate the tunneling
current of the SDQDs as described in Eq. (3). For the complete
equation of motion for Green's functions and correlation
functions, please refer to Appendix A of our previous work
[\onlinecite{DavidK}]. The Green's functions proportional to
$t^2_{LR}/U_{LR}$ are neglected in [\onlinecite{DavidK}], so the
transmission coefficient in Eq. (4) is valid only when
$t_{LR}/U_{LR} < 1$.

\section{Charge stability diagram and electrical conductance for
the case of  $t_{LR} > \Gamma_t$}

\numberwithin{figure}{section}

\numberwithin{equation}{section}

In Fig. 2, we present the charge stability diagrams and tunneling
current under small applied biases ($V_{Bias}= \pm 1$~mV) for the
condition $t_{LR} = \Gamma_t = 1$~meV, where distinguishing
between sequential tunneling and coherent tunneling is
challenging. To better understand the coherent tunneling process,
we consider the condition $t_{LR} > \Gamma_t$. In Fig. C.1, we
show the charge stability diagram ($N_t$) and electrical
conductance ($G_e$) as functions of two gate voltages ($V_{g,L}$
and $V_{g,R}$) at $\Gamma_t = 1$~meV and $T = 12$~K, using the
contour color-fill presentation.

Diagrams (a) and (b) in Fig. C.1 display the calculated values of
$N_t$ and $G_e$ at $t_{LR} = 1$~meV, while diagrams (c) and (d)
show the same quantities for $t_{LR} = 5$~meV. Diagrams (a) and
(b) in Fig. C.1 can be directly compared with diagrams (a) and (b)
in Fig. 2. In Fig. C.1(a), the region marked as (1,1) exhibits a
honeycomb shape, which is consistent with the findings in ref
[\onlinecite{SarmaSD}]. Additionally, the particle number
transition zones are clearly visible between the different
plateaus. In Fig. C.1(b), we observe eight small spots
corresponding to eight configurations for the case of $t_{LR} =
\Gamma_t$. As shown in this panel, charge transport occurs only at
the particle number transition zones. When $t_{LR}
> \Gamma_t$ (coherent tunneling regime), the region marked as ($N_L,N_R$) in
the charge stability diagram undergoes a
modification[\onlinecite{MolitorF}]. In particular, the particle
number transition zones change significantly, as illustrated in
Fig. C.1(c). The electrical conductance exhibits butterfly-like
shapes in Fig. C.1(d). In the symmetric case ($V_{g,L} =
V_{g,R}$), each butterfly wing is symmetrical. In the asymmetric
case ($V_{g,L}\neq V_{g,R}$), the wings of each butterfly become
asymmetrical.

\begin{figure}[h]
\centering
\includegraphics[trim=1.cm 0cm 1.cm 0cm,clip,angle=0,scale=0.3]{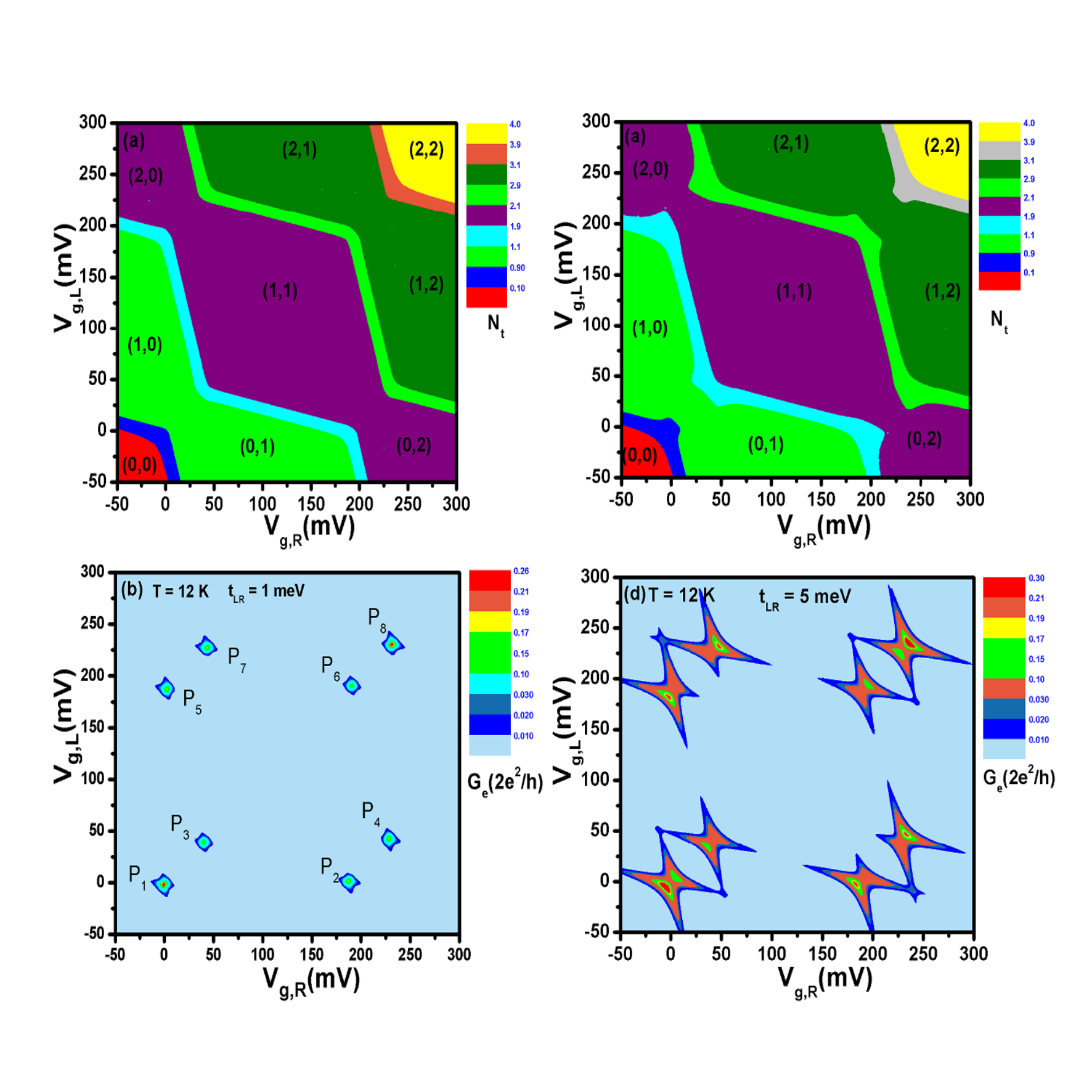}
\caption{(a)Total occupation number ($N_t$) and (b) electrical
conductance ($G_e$) as functions of gate voltages $V_{g,L}$ and
$V_{g,R}$ at a temperature of $T = 12$~K and $t_{LR} = \Gamma_t =
1$~meV. (c) Total occupation number ($N_t$) and (d) electrical
conductance ($G_e$) calculated for $t_{LR} = 5$~meV, with the same
temperature of $T = 12$~K and $\Gamma_t = 1$~meV. All other
physical parameters are the same as those of Fig. 2. The
electrical conductance is expressed in units of $G_0 =
\frac{2e^2}{h}$.}
\end{figure}

%\section{}
%\subsection{Derivation of the tunneling current formula using Dyson's equations\label{App:TC_l} }
\mbox{}\\

%\appendix
%\numberwithin{figure}{section}
%\section{Electronic band structures}

%\numberwithin{equation}{section}

{\bf Data Availability}\\

The data that supports the findings of this study are available
within the article.\\

\textbf{Conflicts of interest }\\

There are no conflicts to declare.\\

{\bf Acknowledgments}\\
We thank Yia-Chung Chang and  Shiue-Yuan Shiau for their valuable
intellectual input. This work was supported by the National
Science and Technology Council, Taiwan under Contract No. MOST
107-2112-M-008-023MY2.

\mbox{}\\
E-mail address: mtkuo@ee.ncu.edu.tw\\

\end{document}